\documentclass[aps,prl,final,twocolumn,letterpaper]{revtex4}
\usepackage{graphicx}
\usepackage{epstopdf}
\usepackage{amsmath}
\usepackage{physics}
\usepackage{bm}
\usepackage{amssymb}
\usepackage{xcolor}
\usepackage{array}
\usepackage{multirow}
\usepackage{lipsum}
\usepackage[T1]{fontenc}
\usepackage{multirow}
\usepackage[english]{babel}
\usepackage{hyperref}

\renewcommand{\v}[1]{{\mathbf{\boldsymbol{#1}}}}

\usepackage{comment}
\usepackage{float}
\usepackage{amsthm}
\usepackage{mathtools}

\begin{document}

\title{Electron bubbles in highly excited states of the lowest Landau level}
\author{David~D.~Dai}
\email{dddai@mit.edu}
\author{Liang~Fu}
\affiliation{Department of Physics, Massachusetts Institute of Technology, Cambridge, Massachusetts 02139, USA}

\date{\today}

\begin{abstract}
We study the entire energy spectrum of an electron droplet in the lowest Landau level. 
By exact diagonalization calculations, we find highly excited states in the middle of the spectrum that display unexpected density distribution and pair correlation.
We show that these exceptional excited states contain tightly bound electron bubbles with local filling $\nu = 1$ that form various ordered structures.
Remarkably, these bubble excited states are shown to exist for both the $1/r$ Coulomb interaction and the $1/r^3$ dipole interaction.  
The experimental realization of bubble excited states in moir\'e materials under a magnetic field is also discussed.
\end{abstract}

\pagestyle{myheadings}
\thispagestyle{empty}
\maketitle

\textit{Introduction---}  
As the prototypical flat topological band, the Landau level is an ideal setting for exploring quantum many-body phenomena. The study of the fractional quantum Hall effect in the lowest Landau level opened a new era in condensed matter physics \cite{qh40yrs} and inspired fundamental research on fractional charge and statistics \cite{Laughlin_PRL, PRL_frac_stat, Haldane_hierarchy, Halperin_hierarchy, QM_fracspin, frac_charge_obs, frac_stat_obs}, composite fermions \cite{CF_PRL, HLR, NPB1991, CF_1stexp, CF_SdH, CF_comens_exp}, and topological order \cite{RMP_topo, RVB_FQH, FQH_GSD}. 
These groundbreaking developments concerned only the ground state and low-lying excitations. 
Highly excited energy eigenstates of a partially filled Landau level have been far less studied. 
The exponentially large density of many-body eigenstates in the middle of the energy spectrum makes it difficult for ordered structures to survive there, so one might expect highly excited states to resemble a thermal liquid.

In this work, we study the entire energy spectrum of an electron droplet in the lowest Landau level. 
By exact diagonalization calculations, we find highly excited states in the middle of the spectrum that display unexpected density distribution and pair correlation. 
These exceptional excited states are identified as electron droplets containing tightly bound electron bubbles with local filling $\nu = 1$, which are formed by the Coulomb interaction in a flat band. 
Remarkably, these bubble excited states are shown to exist for both the $1/r$ Coulomb interaction and the $1/r^3$ dipole interaction.  
The experimental realization of bubble excited states in moir\'e atom arrays under a magnetic field is also discussed.
Our work therefore uncovers highly excited states with charge order in the lowest Landau level and suggests their existence at finite energy density in the thermodynamic limit.

Our study is motivated in part by previous work on higher Landau levels at partial filling, where a rich variety of charge-ordered ground states has been found both theoretically and experimentally \cite{bubbletheory1, bubbletheory2, highLL_exact, bubblenumerics, 2LL_DMRG, stripe_exp1, stripe_exp2, bubble_exp1, bubble_exp2}. These include stripe phases of alternating $\nu = n$ and $\nu = n + 1$ domains ($n$ is the number of fully filled Landau levels), as well as bubble phases where tightly packed electron bubbles form a lattice. In contrast, the ground states of the first and second lowest Landau levels are known to be fractional quantum Hall liquids instead of charge-ordered states \cite{FQH_vs_CDW, stripe_exp1, stripe_exp2, half_highLL_numerics, 1LL_exp, 1LL_nonconventional, bubblenumerics}. However, our work reveals a plethora of bubble states that survive in the lowest Landau level as highly excited states.

\textit{Exact Diagonalization---} The position-space Hamiltonian for a droplet of $N$ electrons confined to a plane and subject to a perpendicular magnetic field is:
\begin{equation}\label{eq:pos_H}
    H = \sum_{i=1}^N \frac{1}{2m}\left[\v p_i - q \v A(\v r_i)\right]^2 + \sum_{i < j} v\left(|\v r_i - \v r_j|\right),
\end{equation}
where $\v r_i$ and $\v p_i$ are the position and canonical momentum respectively of the $i$th electron, $q$ is the electron charge, $\v A(\v r) = \v B \cross \v r / 2$ is the symmetric gauge vector potential, and $v(r)$ is the electron-electron interaction. We work in units where $\hbar = 1$, $l_B = 1/\sqrt{qB} = 1$, and $q > 0$, and we also define the complex coordinates $z_i = x_i + i y_i, \bar{z}_i = x_i - i y_i$. Throughout, we restrict ourselves to the lowest Landau level because we are interested in the limit of a strong magnetic field.

One conserved quantity is the total angular momentum
\begin{equation}\label{eq:L_tot_def}
    L_\text{tot} = \sum_{i=1}^N \left[z_i \partial_i - \bar{z}_i \bar{\partial}_i\right],
\end{equation}
where $\partial_i$ $\left(\bar{\partial}_i\right)$ is the derivative with respect to $z_i$ $\left(\bar{z}_i\right)$, treating $z_i$ and $\bar{z}_i$ as independent. We note that a parabolic confining potential can be added to the Hamiltonian to model a quantum dot in a magnetic field. Since this term is proportional to the total angular momentum within the lowest Landau level, it does not change the eigenstates within a fixed total angular momentum sector.

Because the interaction only depends on the electrons' relative distance but not on the center-of-mass position, the angular momentum of the center of mass
\begin{equation}
    L_\text{CM} = z_\text{CM}\pdv{}{z_\text{CM}} - \bar{z}_\text{CM}\pdv{}{\bar{z}_\text{CM}}, \quad z_\text{CM} = \frac{1}{N}\sum_i z_i
\end{equation}
is also a conserved quantity \cite{girvin_cluster}. The droplet's internal structure depends solely on the angular momentum of the electrons about their center of mass, i.e. $L_\text{internal} = L_\text{tot} - L_\text{CM}$; the center of mass's angular momentum about the origin is irrelevant. To eliminate this redundancy, we always consider sectors where $L_\text{CM} = 0$.

\begin{figure}[t!]
  \centering
  \includegraphics[width=0.47\textwidth]{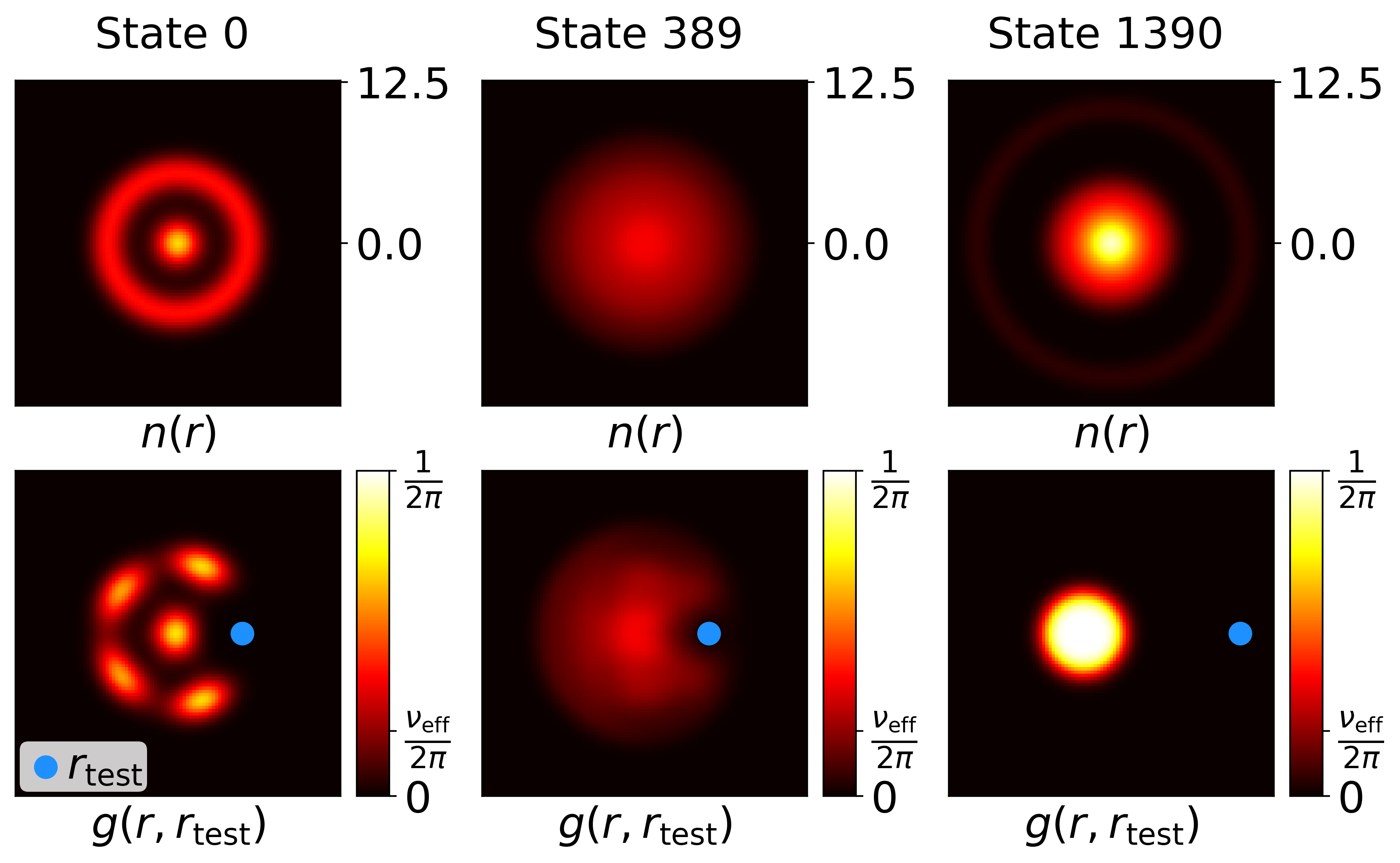}
  \caption{Total electron density and pair correlation function for the ground state, a typical mid-spectrum eigenstate, and the highest-energy state of the droplet $(N, L_\text{tot}, L_\text{CM}, v(r)) = (6, 75, 0, 1/r)$, which has dimension $1391$.}
  \label{fig:gs_therm_he}
\end{figure}

Then the system's symmetry sector is fully specified by the particle number $N$ and total($=$ internal) angular momentum $L_\text{tot}$, which acts as a proxy for the filling fraction. By comparing to the Laughlin $\nu = 1 / m$ state, which has total angular momentum $L_\text{tot} = mN(N-1)/2$, we can see that, in general, the effective filling fraction is $\nu_\text{eff} = N(N - 1)/2L_\text{tot}$ \cite{Laughlin_PRL, girvin_cluster}. 

To probe the average orbital occupancy of the eigenstates, we compute the one-particle reduced density matrix (diagonal due to angular momentum conservation) $\rho^{(1)}_l = \expval{c^\dagger_l c_l}$, where $c_l$ is the annihilation operator for the $l$th symmetric gauge orbital $\phi_l$. $\phi_l$ is concentrated on a ring of radius $r_l \approx \sqrt{2l}$, so $\rho^{(1)}_l$ is also a proxy for the density profile. To probe the internal structure of the eigenstates, we calculate the pair correlation function
\begin{equation}\label{eq:pair_correlation}
    g(\v r, \v r_\text{test}) = \frac{\expval{\mathrm{:}n(\v r)n(\v r_\text{test})\mathrm{:}}}{\expval{n(\v r_\text{test})}},
\end{equation}
where $n(\v r) = \sum_{i=1}^{N} \delta(\v r - \v r_i)$ is the density operator. Using the above normalization, one may show that $\int \text{d}^2 \v r \left[g(\v r, \v r_\text{test})\right] = N - 1$, so $g(\v r, \v r_\text{test})$ may be interpreted as the electron density at $\v r$ conditioned on one of the electrons being clamped at $\v r_\text{test}$. Finally, we also calculate the radial structure factor
\begin{equation}\label{eq:radial_Sl}
    S_l = \expval{\sum_{i < j}\mathcal{P}_l{(ij)}},
\end{equation}
where $\mathcal{P}_l(ij)$ is the projector onto the relative angular momentum $l$ channel for electrons $i$ and $j$. $S_l$ measures the expected number of electron pairs in the relative angular momentum $l$ channel. See the SM for additional details on implementing exact diagonalization \cite{supplement}.

\begin{figure}[t!]
  \centering
  \includegraphics[width=0.47\textwidth]{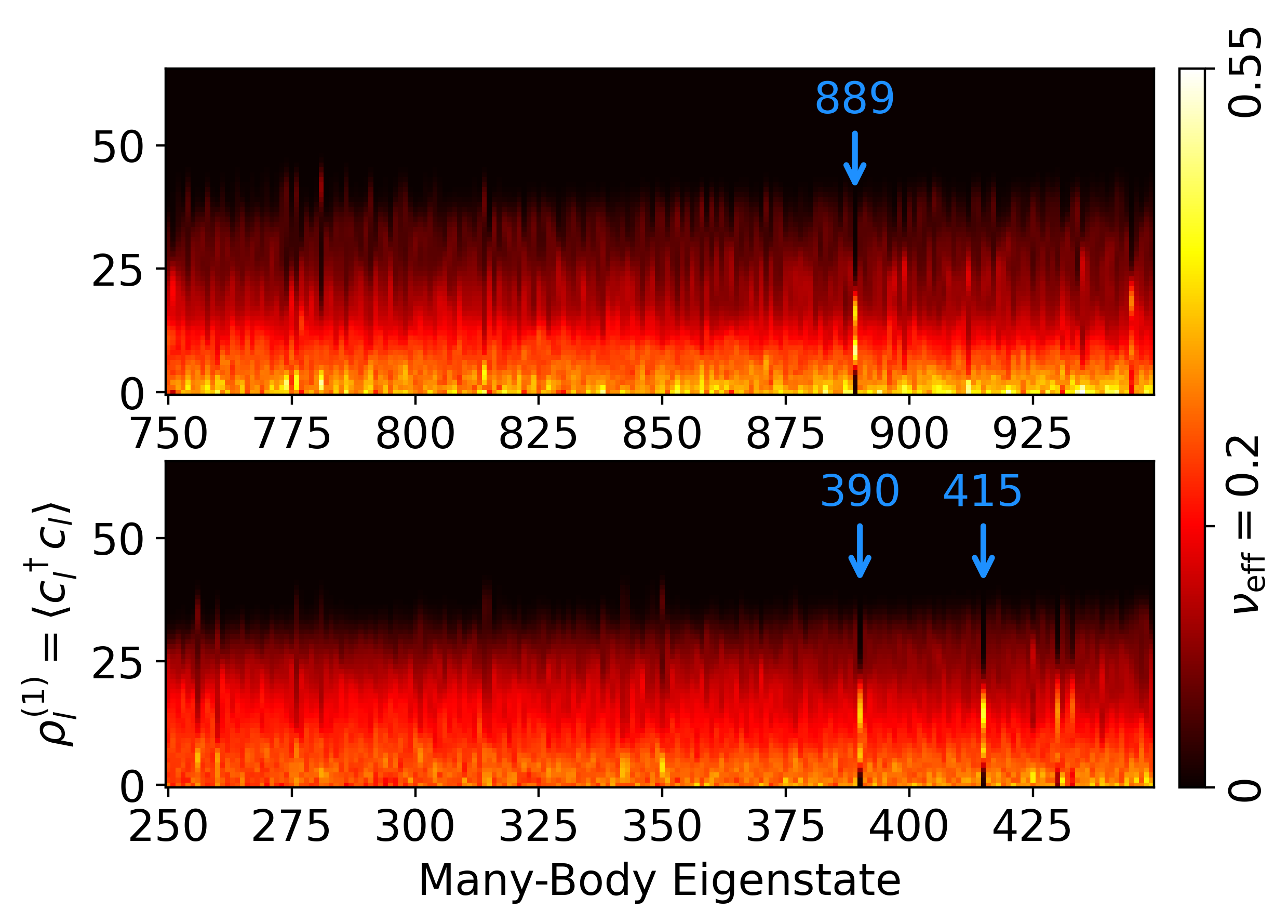}
  \caption{One-particle reduced density matrix $\rho^{(1)}_l = \expval{c_l^\dagger c_l}$ for select eigenstates of the droplet $(N, L_\text{tot}, L_\text{CM}, v(r)) = (6, 75, 0, 1/r)$, where $c_l$ is the annihilation operator for the symmetric gauge orbital $\phi_l$.}
  \label{fig:1p-rdm}
\end{figure}

\textit{ Bubble States---} We now present exact diagonalization results for the droplet $(N, L_\text{tot}, L_\text{CM}, v(r)) = (6, 75, 0, 1/r)$, which models six electrons at effective filling $\nu = 1 / 5$ and has a Hilbert space dimension of $1391$.

Fig. \ref{fig:gs_therm_he} shows the electron density and pair correlation function for the ground state, a typical mid-spectrum eigenstate, and the highest-energy state. The ground state is a  Wigner molecule \footnote{Although the ground state of the lowest Landau level at $\nu=1/5$ filling is a fractional quantum Hall liquid in the thermodynamic limit, phases of matter are not well-defined for a finite-size system. We consider the ground state of the $N=6$ disk system to be a Wigner molecule based on the clear features in its correlation functions.} with one electron localized at the origin and the other five localized on a regular pentagon, which best minimizes the potential energy. The highest-energy state consists of a bubble of $N-1=5$ electrons of local filling fraction $\nu = 1$ and one electron orbiting the bubble at long distance. The formation of a $\nu=1$ bubble maximizes the interaction energy. The ejection of one of the electrons from the $\nu = 1$ bubble is necessary to achieve the desired total angular momentum. This assignment is further supported by $1391$'s structure factor $S_1 = 5.0780$, almost exactly the same as the $N=5,\nu=1$ state's $S_1 = 5.0781$. Meanwhile, the mid-spectrum eigenstate's density profile and pair correlation are essentially featureless, consistent with it being a thermal liquid (see SM for additional featureless states \cite{supplement}).

\begin{figure}[t!]
  \centering
  \includegraphics[width=0.47\textwidth]{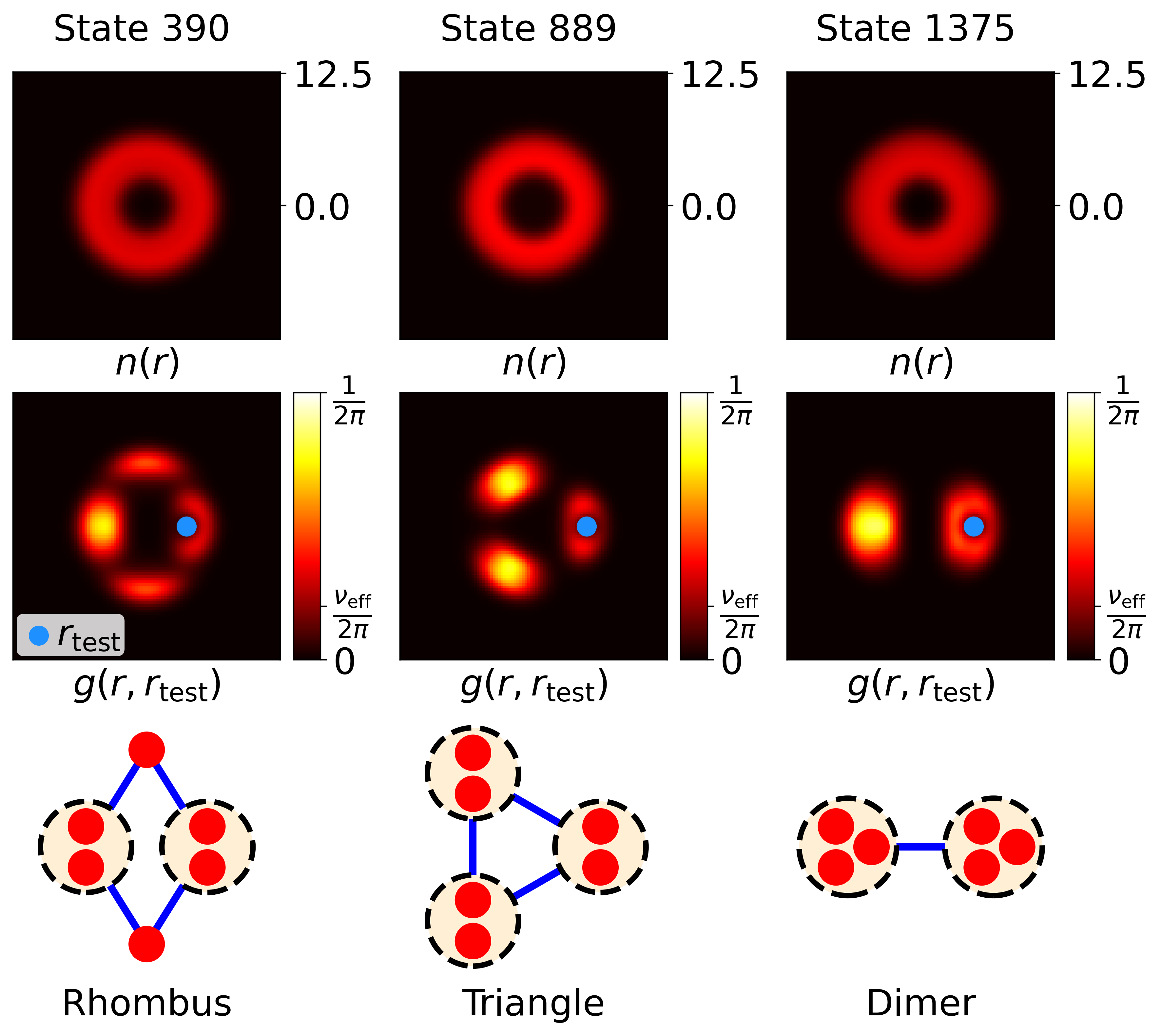}
  \caption{Total electron density, pair correlation function, and schematic for eigenstates $390$, $889$, and $1375$ in the cluster $(N, L_\text{tot}, L_\text{CM}, v(r)) = (6, 75, 0, 1/r)$.}
  \label{fig:390_889_1375}
\end{figure}

To search for interesting eigenstates, we examine the one-particle reduced density matrices $\rho^{(1)}_l$ ($=$average orbital occupancies) of the eigenstates, some of which are plotted in Fig. \ref{fig:1p-rdm}. A priori, one expects that away from the spectrum's edges, the orbital occupancies will change smoothly with the eigenstate index, and mid-spectrum states will not have any notable features in their density profile or correlation functions. Although we observe this for the most part, such as for state $389$ in Fig. \ref{fig:gs_therm_he}, there are exceptional many-body eigenstates such as $390$, $415$, and $889$ with $\rho^{(1)}_l$ tightly concentrated on a narrow band of $l$, which is very different from most other states. To further understand these special eigenstates, we calculate the pair correlation functions of representative examples $390$, $889$, and $1375$, which are shown in Fig. \ref{fig:390_889_1375}. $1375$ is very close to the top of the spectrum, but we have still included it because it is qualitatively different from the other highest-energy states.

We start by analyzing $390$'s pair correlation function. When $\v r_\text{test}$ is placed near the inner boundary of the ring where the density is concentrated, the region where another electron is likely to be found is composed of four distinct lobes. One lobe surrounds $\v r_\text{test}$ and contains close to one electron (these populations are computed by integrating $g(\v r, \v r_\text{test})$ over $\v r$ inside the lobe --- see SM for additional figures \cite{supplement}). Another lobe is diametrically opposed to $\v r_\text{test}$ and contains two electrons. The last two lobes are $90^{\circ}$ clockwise and counterclockwise of $\v r_\text{test}$ and contain one electron each. If $\v r_\text{test}$ is placed near the outer boundary of the ring, there are three lobes, one diametrically opposed to $\v r_\text{test}$ containing approximately one electron, and two $90^{\circ}$ clockwise and counterclockwise of $\v r_\text{test}$ containing approximately two electrons each.

This pair correlation shows that eigenstate $390$ is composed of two bubbles, each containing two electrons bound in a $\nu = 1$ state, plus two lone electrons. The high magnetic field stabilizes the bubbles by forcing the constituent electrons to orbit around each other, creating a bound state despite the repulsive interaction \cite{laughlin_3body, girvin_cluster}. Each bubble acts as a composite particle, allowing us to create a rigid orbit (an orbit where all components rotate at the same angular velocity) by placing the bubbles on the vertices of a rhombus. Finally, the quantum superposition over rotations of the rhombus yields a rotationally symmetric eigenstate. The existence of electron bubbles is further supported by eigenstate $390$'s structure factor $S_1 = 1.9681$, indicating that there are nearly two pairs of electrons in the $l=1$ relative angular momentum channel.

States $889$ and $1375$ have similar explanations for their correlation functions. $889$ consists of three two-electron $\nu = 1$ bubbles located on the vertices of an  equilateral triangle. This is also supported by $889$'s structure factor $S_1 = 2.9273$, indicating that nearly three pairs of electrons are in the $l=1$ channel. $1375$ is a  dimer of two three-electron $\nu = 1$ bubbles. This is also supported by $1375$'s structure factor $S_1 = 4.4902$, which is close to twice that of the $N=3, \nu = 1$ state, which has $S_1 = 2.25$.

\begin{figure}[t!]
  \centering
  \includegraphics[height=0.41\textwidth]{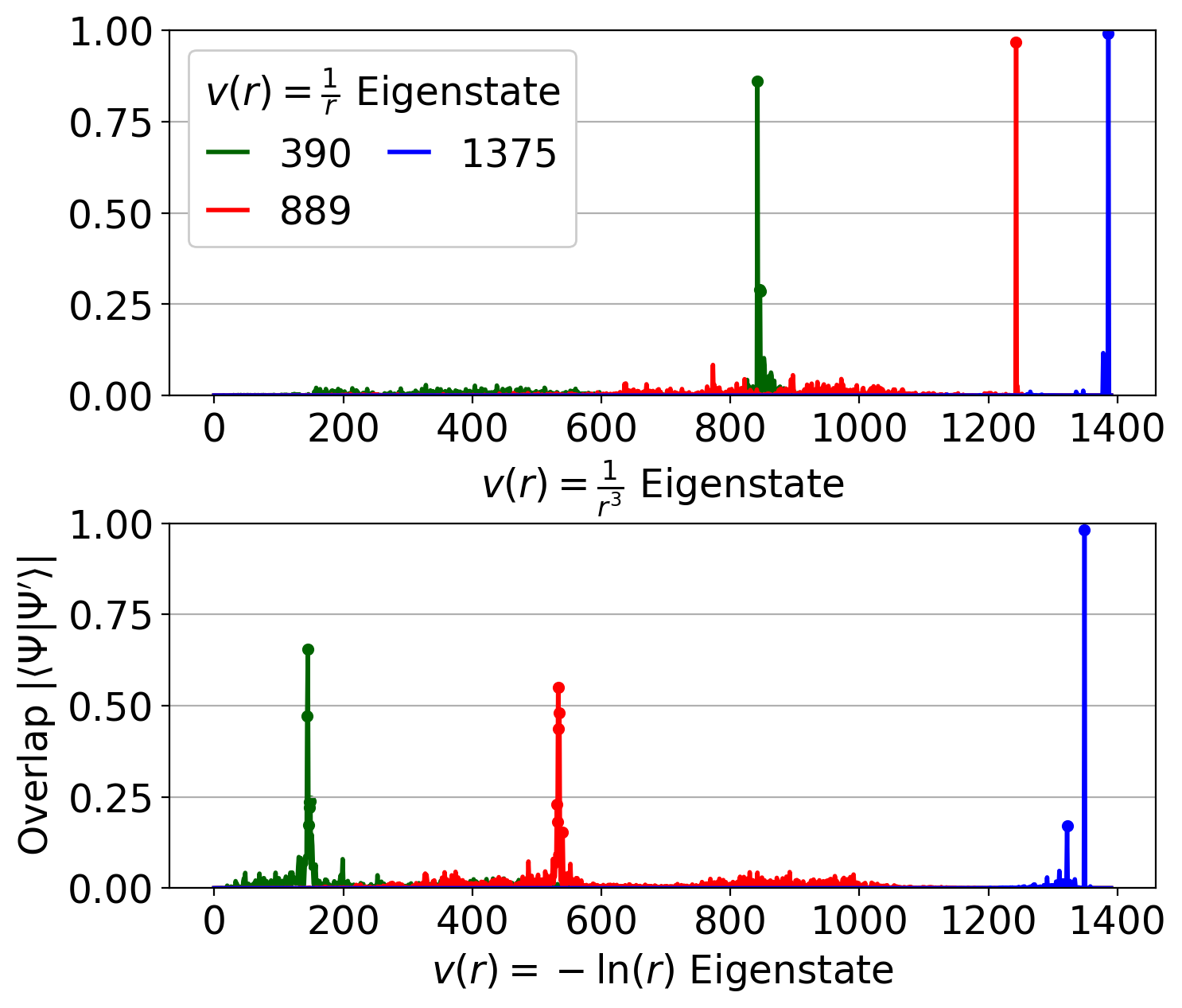}
  \caption{Overlaps $|\bra{\Psi}\ket{\Psi'}|$ of the entire $v(r) = 1/r^3$ and $v(r) = -\ln(r)$ spectra with eigenstates $390$, $889$, and $1375$ of the $v(r) = 1/r$ spectrum.}
  \label{fig:overlaps}
\end{figure}

\textit{Robustness of bubbles---} To assess the robustness of electron bubbles in highly excited states of the lowest Landau level, we have also performed exact diagonalization for the same symmetry sector $(N, L_\text{tot}, L_\text{CM}) = (6, 75, 0)$ but different interactions $v(r) = 1/r^3$ and $-\ln(r)$. Fig. \ref{fig:overlaps} plots the overlaps $|\bra{\Psi}\ket{\Psi'}|$ of the entire $1/r^3$ and $-\ln(r)$ spectra with the $1/r$ states $390$, $889$, and $1375$.

The $1/r^3$ spectrum contains almost identical triangle and dimer bubble states. Eigenstates $842_{1/r^3}$, $1243_{1/r^3}$, and $1386_{1/r^3}$ (here $N_{v}$ denotes the $N$th eigenstate for interaction $v$) have overlaps $0.8617$, $0.9683$, and $0.9929$ with the previously analyzed $390_{1/r}$, $889_{1/r}$ and $1375_{1/r}$ respectively. Furthermore, the pair correlation functions of the $1/r^3$ states have the same features as their $1/r$ partners (see SM for additional plots \cite{supplement}).

This universality across interactions is remarkable given the lack of an energy gap protecting these excited states, but it can be understood from the structure of the bubble states. The compactness of the $\nu = 1$ state prevents changes in each bubble's internal structure, and the geometric arrangement of the bubbles within the electron droplet is tightly constrained by symmetry. For example, $889_{1/r}$'s equilateral triangle and $1375_{1/r}$'s dimer are completely fixed by symmetry, while $390_{1/r}$'s rhombus only has one free parameter, the angle of the rhombus. The rhombus not being fully constrained by symmetry is consistent with the lower overlap between the $1/r^3$ and $1/r$ rhombus states.

On the other hand, the $-\ln(r)$ spectrum contains a state very close to $1375_{1/r}$, but the rhombus and triangle states are suppressed. Analogous states can still be found in the $-\ln(r)$ spectrum, but the features in their correlation functions are much fainter (see SM \cite{supplement}). These results are explained by $-\ln(r)$'s slower fall-off with distance. For the bubble states to be good energy eigenstates, there must be a separation of energy scales between intra and inter-bubble interactions. An electron-electron interaction that falls off too slowly with distance couples the intra-bubble degrees of freedom and the bubbles' centers of mass. This disrupts and eventually breaks apart the bubbles. For $v(r) = -\ln(r)$, the rhombus and triangle are significantly weakened by this effect. However, the two three-electron bubbles in the dimer are presumably far enough apart for the state to be nearly unchanged.

\textit{Conclusion--- } Our work initiates the study of highly-excited bubble states in the lowest Landau level. 
Exact diagonalization calculations find eigenstates in the middle of the spectrum with unexpected density distribution and pair correlation function. 
We show that these exceptional highly excited states contain bubbles of tightly bound electrons in a local $\nu=1$ state, similar to the ground states of partially filled higher Landau levels. 
Additional exact diagonalization calculations show that very similar bubble states exist for both the $1/r$ Coulomb interaction and $1/r^3$ dipole interaction.

The presence of a small number of eigenstates in the middle of the spectrum with exceptional pair correlation and density profile, combined with the remaining eigenstates being essentially featureless, is highly reminiscent of many-body scarring. Recently discovered in spin models \cite{scar_first_exp, PXP_theory}, many-body scars are exceptional eigenstates that have finite energy density yet have ground state-like properties such as area-law entanglement \cite{scar_review, PXP_theory} in defiance of the eigenstate thermalization hypothesis \cite{ETH_1, ETH_2, ETH_3}. Scars have been found in a variety of models such as the PXP spin chain (modeling Rydberg atom arrays) \cite{scar_first_exp, PXP_theory, scar_review}, spin-$1$ AKLT model \cite{AKLT_exact, AKLT_entanglement}, XY models \cite{1xy_scar}, Hubbard models \cite{hubbard-eta}, and partially-filled Landau levels on a thin-torus (which map to spin chains) \cite{LL_thin_torus}.

In contrast, our work finds exceptional excited states of electrons in a continuum system, albeit with a finite number of electrons. We speculate that electron bubbles survive in the thermodynamic limit and may form a lattice, giving  rise to quantum many-body scars. This is plausible for the same reason that our 6-electron bubble states are so robust against variations in the interaction: the compactness of the $\nu = 1$ state prevents changes in each bubbles' internal structure, and the lattice symmetry renders the net force on each bubble zero. We leave the fate of the bubbles in the thermodynamic limit and other compelling questions to future studies. 

It will be interesting to experimentally probe highly excited states in a Landau level droplet and search for signatures of electron bubbles. We propose that a periodic array of multi-electron molecules can be formed in semiconductor moir\'e superlattices, in which each molecule is locked to a moir\'e potential minimum as experimentally observed \cite{moire_atom_array, wigner_molecule}. Applying a magnetic field may produce an array of multi-electron Landau level droplets \cite{hofstader}. We also envision that pumping this system with a light pulse and probing its long-lived response \cite{corr_insulator, corr_insulator_2} may reveal spectroscopic signatures of electron bubbles such as their vibrational modes.     
In a broader context, we hope that our work stimulates both the theoretical and experimental investigation of highly-excited non-thermal states in realistic condensed matter settings such as flat band systems.

\textit{Acknowledgements---} We are grateful to Philip Crowley, Margarita Davydova, Aidan Reddy, Ahmed Abouelkomsan, and Max Geier for helpful discussions. This work was supported by the Simons Investigator Award from the Simons Foundation. DDD was supported by the Undergraduate Research Opportunities Program at MIT. The authors acknowledge the MIT SuperCloud and Lincoln Laboratory Supercomputing Center for providing high-performance computing resources.

\bibliography{references}

%%%%%%%%%%%%%%%%%%%%%%%%%%%%%%
%%% Supplementary Material %%%
%%%%%%%%%%%%%%%%%%%%%%%%%%%%%%

% Supplementary will be exported to extra file

\clearpage

% add a big title
\onecolumngrid
\begin{center}
\textbf{\large Electron bubbles in highly excited states of the lowest Landau level\\Supplemental Material}\\[0.5cm]
David D. Dai and Liang Fu\\%[0.15cm]
{\itshape{\small Department of Physics, Massachusetts Institute of Technology, Cambridge, Massachusetts 02139, USA}}\\
\end{center}
\twocolumngrid

% reset equation counters
\setcounter{page}{1}
\setcounter{equation}{0}
\setcounter{figure}{0}
\setcounter{table}{0}
\setcounter{section}{0}

% add prefix "S" to all equations
\renewcommand{\v}[1]{{\mathbf{\boldsymbol{#1}}}}
\renewcommand{\theequation}{S\arabic{equation}}
\renewcommand{\thefigure}{S\arabic{figure}}
\renewcommand{\thetable}{S\arabic{table}}

\begin{widetext}
\tableofcontents{}
\section{Additional Figures}

\begin{figure*}[h!]
  \centering
  \includegraphics[width=0.96\textwidth]{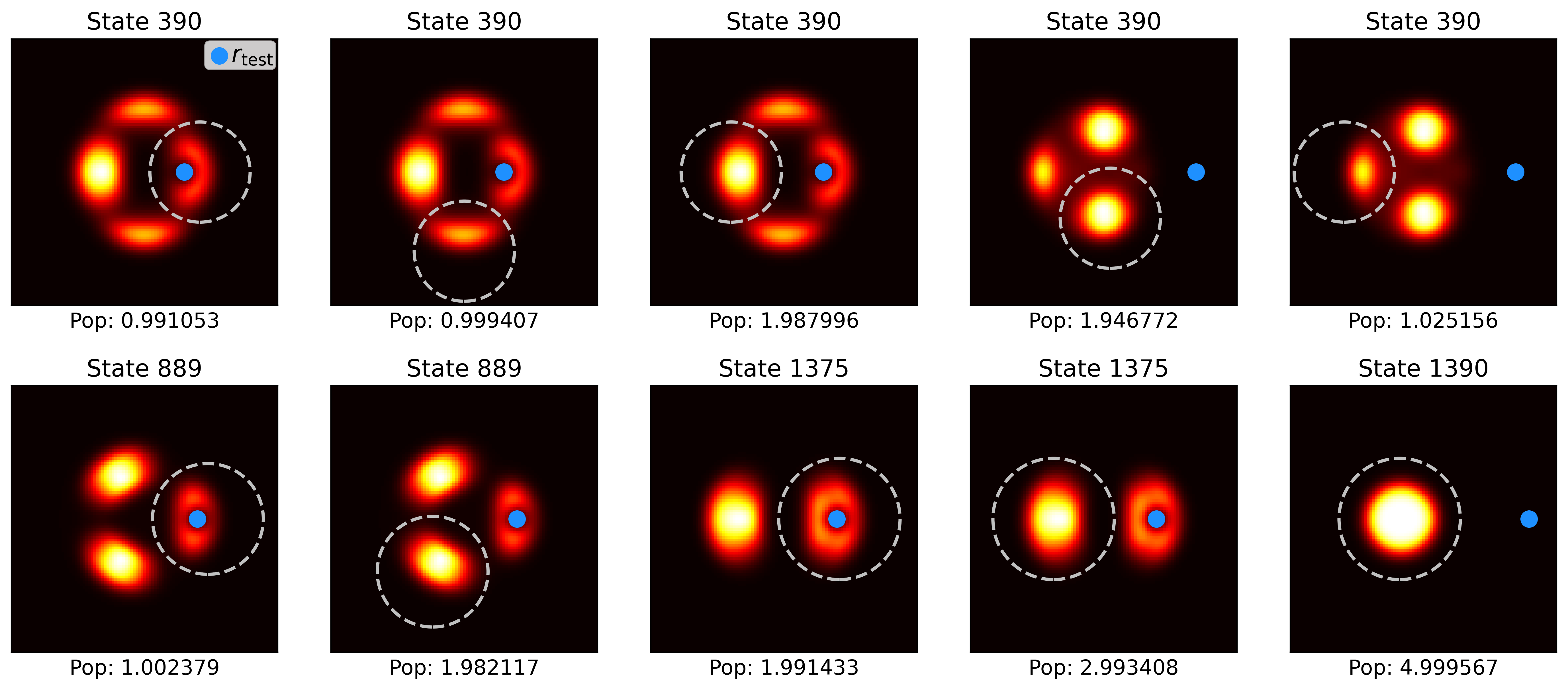}
  \caption{Droplet $(N, L_\text{tot}, L_\text{CM}, v(r)) = (6, 75, 0, 1/r)$. Detailed population analysis of pair correlation functions for bubble states $390$, $889$, and $1375$. The correlation lobes contain close to integer numbers of electrons (the population reported is that inside the circled region).}
  \label{fig:pop_analysis}
\end{figure*}

\begin{figure*}[h!]
  \centering
  \includegraphics[height=0.42\textheight]{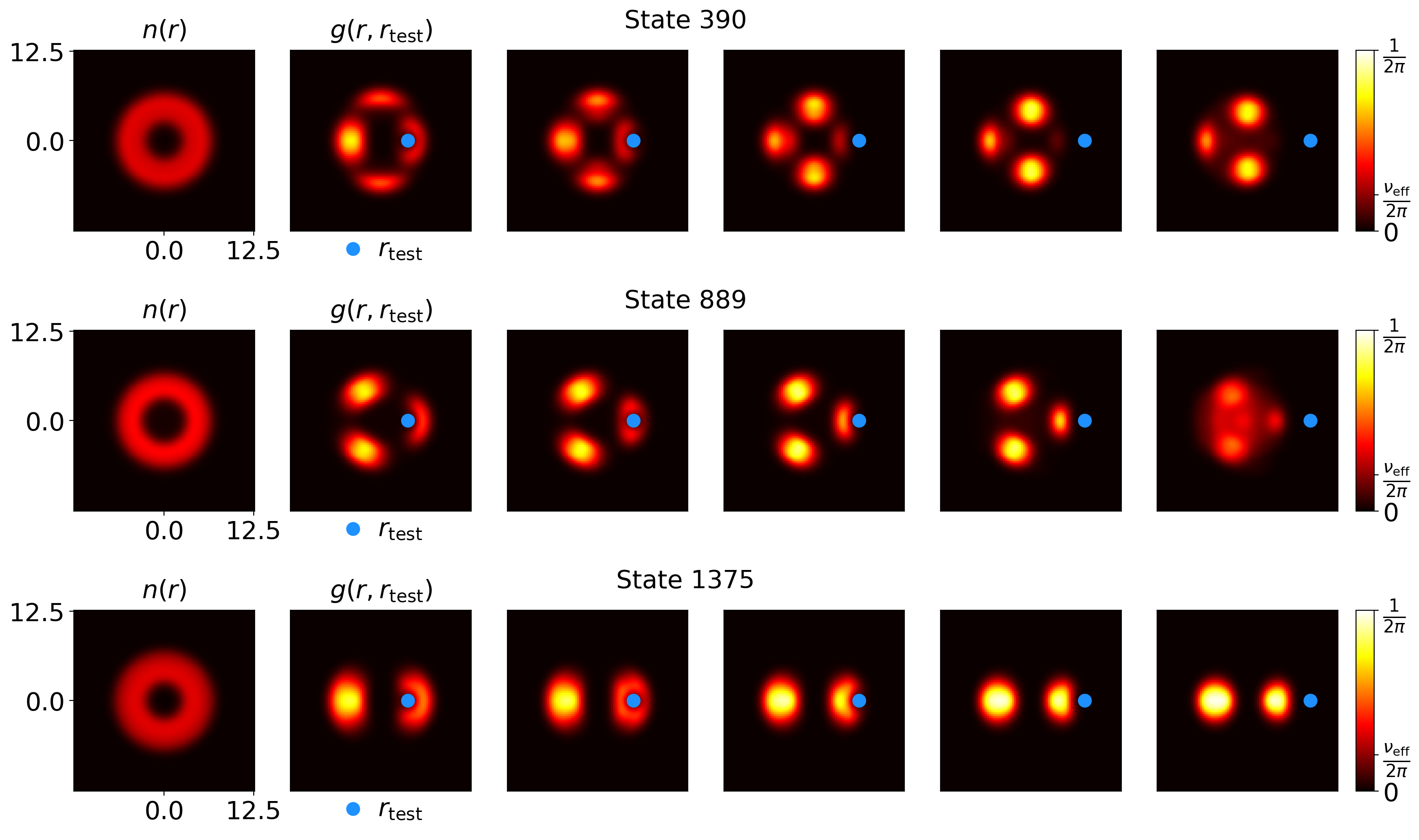}
  \caption{Droplet $(N, L_\text{tot}, L_\text{CM}, v(r)) = (6, 75, 0, 1/r)$. Additional $r_\text{test}$ locations for the correlation functions of bubble states $390$, $889$, and $1375$.}
  \label{fig:coulomb_bubble}
\end{figure*}

\begin{figure*}[h!]
  \centering
  \includegraphics[height=0.42\textheight]{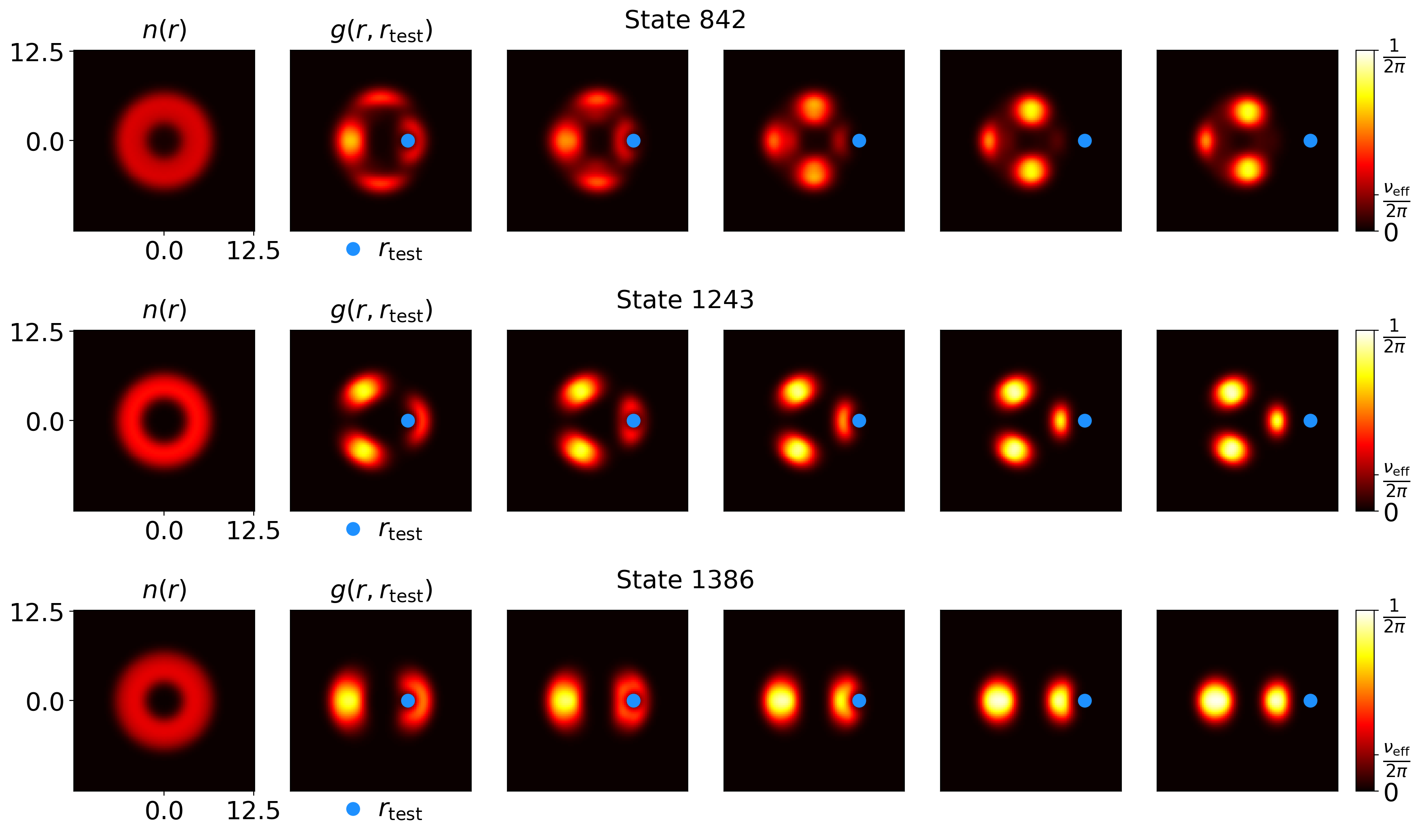}
  \caption{Droplet $(N, L_\text{tot}, L_\text{CM}, v(r)) = (6, 75, 0, 1/r^3)$. Pair correlation functions for the dipole bubble states, which are almost indistinguishable from the Coulomb bubble states.}
  \label{fig:dipole_bubble}
\end{figure*}

\begin{figure*}[h!]
  \centering
  \includegraphics[height=0.42\textheight]{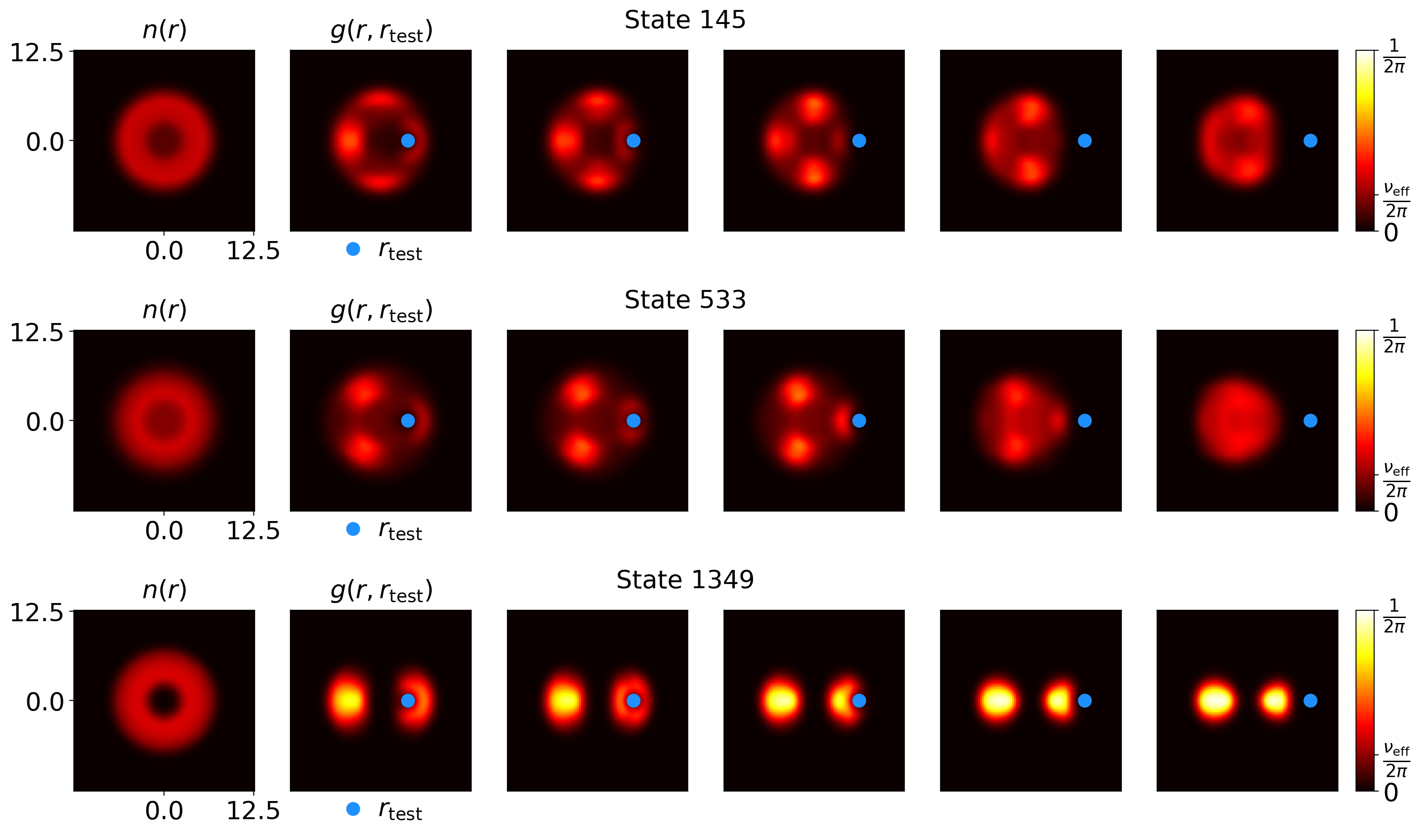}
  \caption{Droplet $(N, L_\text{tot}, L_\text{CM}, v(r)) = (6, 75, 0, -\ln(r))$. Pair correlation functions for the $v(r) = -\ln(r)$ eigenstates that had the highest overlap with the $v(r)=1/r$ bubble states. States $145_{-\ln(r)}$ and $533_{-\ln(r)}$ have much fainter features than their Coulomb counterparts, but clear rhombus and triangle-like patterns are still visible. Meanwhile, the bubble dimers $1349_{-\ln(r)}$ and $1375_{1/r}$ are very similar.}
  \label{fig:log_bubble}
\end{figure*}

\begin{figure*}[h!]
  \centering
  \includegraphics[height=0.42\textheight]{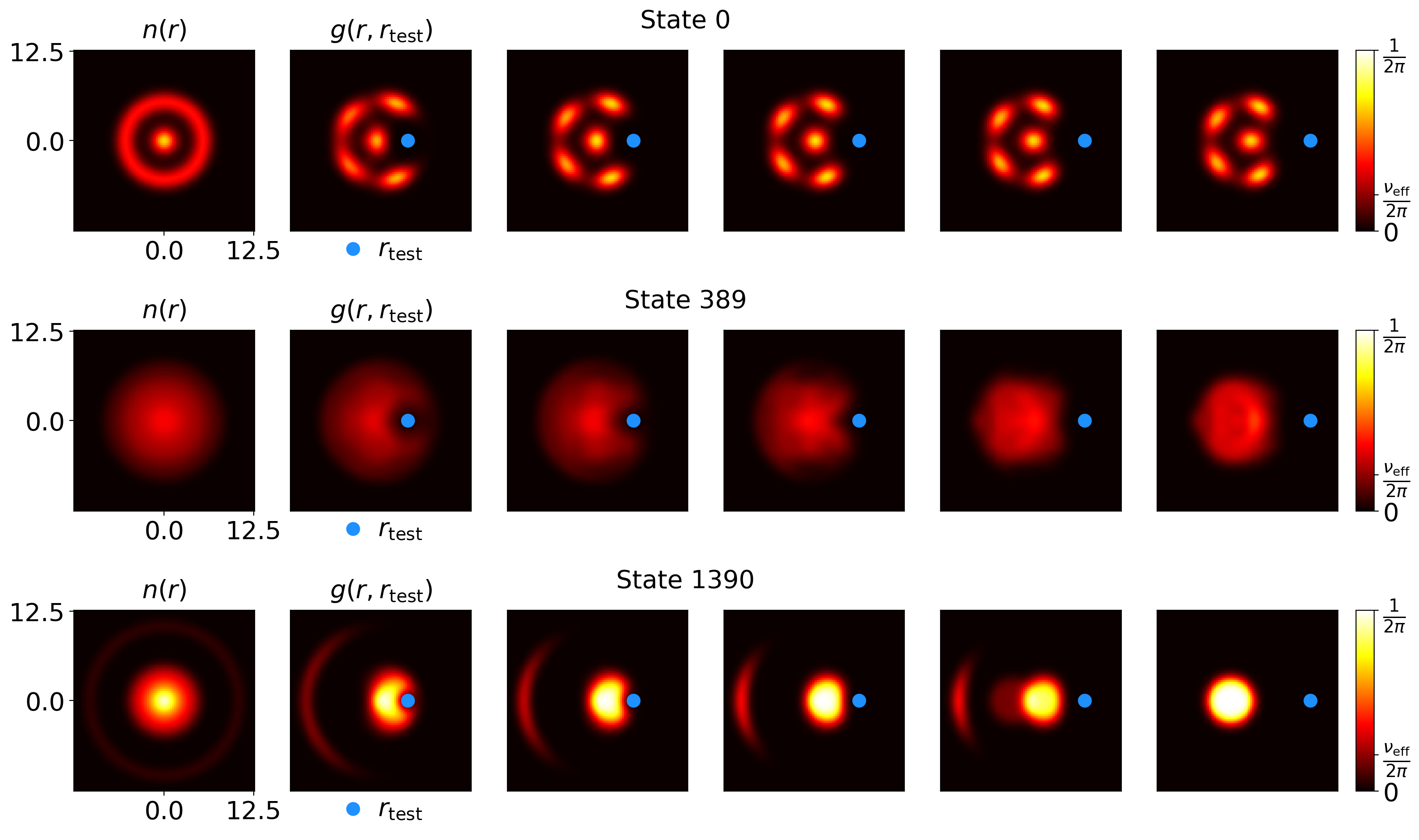}
  \caption{Droplet $(N, L_\text{tot}, L_\text{CM}, v(r)) = (6, 75, 0, 1/r)$. Additional $r_\text{test}$ locations for the correlation functions of the ground state, a typical mid-spectrum eigenstate $389$, and the highest-energy state $1390$.}
  \label{fig:more_ground}
\end{figure*}

\begin{figure*}[h!]
  \centering
  \includegraphics[height=0.42\textheight]{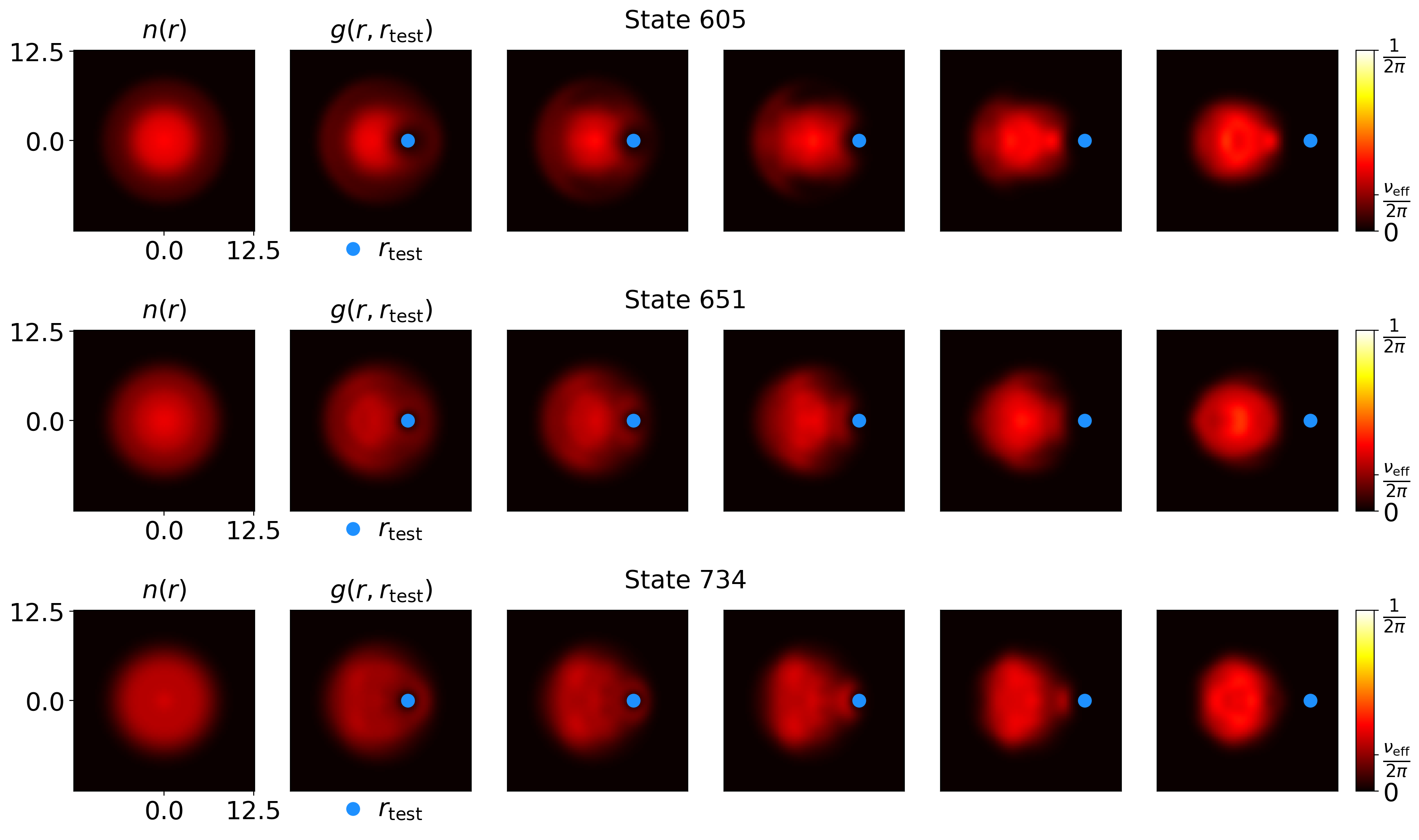}
  \caption{Droplet $(N, L_\text{tot}, L_\text{CM}, v(r)) = (6, 75, 0, 1/r)$. Additional featureless mid-spectrum states.}
  \label{fig:more_boring}
\end{figure*}
\end{widetext}

\clearpage
\newpage
\section{Landau Levels in the Disk Geometry}
\subsection{Landau Level Raising and Lowering Operators}

The position space Hamiltonian for $N$ electrons confined to an infinite plane and subject to a perpendicular magnetic field is:
\begin{equation}\label{eq:SM_pos_H}
    H = \sum_{i=1}^N \frac{1}{2m}\left[\v p_i - q \v A(\v r_i)\right]^2 + \sum_{i < j} v\left(|\v r_i - \v r_j|\right),
\end{equation}
where $\v r_i$ and $\v p_i$ are the position and canonical momenta respectively of the $i$th electron, $\hbar = 1$, $q$ is the electron charge, and $\v A(\v r)$ is a vector potential generating a uniform perpendicular magnetic field. Throughout, we use the convention that Latin letters $i,j\ldots$ index particles, Greek letters $\mu,\nu \ldots$ index the in-plane direction, and $\epsilon_{\mu\nu}$ is the Levi-Civita symbol in two dimensions. Additionally, we use the convention that $q$ is positive, which is more natural and allows us to work with the complex coordinate $z = x + iy$ instead of $z = x - iy$ later.

The mechanical momentum (here, we drop the index $i$ and work with only one particle, as the generalization to many particles is obvious) is:
\begin{equation}\label{eq:mechanical_momenta}
    \v \pi = \v p - q \v A(\v r).
\end{equation}
The mechanical momentum's components satisfy the commutator:
\begin{equation}\label{eq:mechanical_commutator}
    \comm{\pi_\mu}{\pi_\nu} = \frac{i}{l_B^2} \epsilon_{\mu\nu},
\end{equation}
where the magnetic length is defined as $l_B = 1/\sqrt{qB}$. The mechanical momentum allows us to define the Landau level raising and lowering operators:
\begin{equation}\label{eq:LL_ladder}
\begin{aligned}
    a &= \frac{l_B}{\sqrt{2}}\left(\pi_x + i\pi_y\right),\\
    a^\dagger &= \frac{l_B}{\sqrt{2}}\left(\pi_x - i\pi_y\right),
\end{aligned}
\end{equation}
which satisfy:
\begin{equation}
\begin{gathered}
    \comm{a}{a^\dagger} = 1, \\
    H = \frac{1}{2m} (\v p - q\v A)^2 = \omega_B \left(a^\dagger a + \frac{1}{2}\right).
\end{gathered}
\end{equation}
Above, we have defined the cyclotron frequency $\omega_B = qB/m$, which is the energy scale of the magnetic field. For weak interactions $V \ll \omega_B$, the majority of the physics occurs in the lowest Landau level where $a\ket{\psi} = 0$.

\subsection{Guiding Center Operators}

Classically, a particle in a magnetic field follows a circular trajectory of angular frequency $\omega_B$ around a fixed guiding center. The particle's energy comes from the rotational kinetic energy around the guiding center, which depends on the orbital radius. Meanwhile, the Landau level's degeneracy comes from the fact that there are many points around which the electron could orbit. In the quantum theory, the Landau level raising and lowering operators change the orbital radius and hence the energy. Analogously, we should be able to identify operators in the quantum theory that correspond to the guiding center, enabling us to move within the highly-degenerate Landau levels.

From the Lorentz force law, it is easy to see that a particle's displacement from the guiding center is locked to its momentum:
\begin{equation}\label{eq:guiding_center}
    R_\mu = r_\mu + l_B^2 \epsilon_{\mu\nu} \pi_\nu.
\end{equation}
Above, we use $\v \pi$ as it and not $\v p$ corresponds to the physical velocity of the electron. The guiding center commutes with the mechanical momenta and thus the kinetic energy:
\begin{equation}\label{eq:guide_p_comm}
\begin{aligned}
    \comm{\pi_\mu}{R_\nu} &= \comm{\pi_\mu}{r_\nu + l_B^2 \epsilon_{\nu \kappa}\pi_\kappa}\\
    &= -i\delta_{\mu\nu} + l_B^2 \epsilon_{\nu\kappa} \frac{i}{l_B^2} \epsilon_{\mu\kappa}\\
    &= -i\delta_{\mu\nu} + i\delta_{\mu\nu}\\
    &=0,
\end{aligned}
\end{equation}
as expected from our classical intuition. However, the guiding center's $x$-coordinate and $y$-coordinate do not commute:
\begin{equation}\label{eq:R_x_R_y_comm}
\begin{aligned}
    \comm{R_x}{R_y} &= \comm{x + l_B^2 \pi_y}{y - l_B^2 \pi_x}\\
    &= -il_B^2 - il_B^2 + l_B^4 \frac{i}{l_B^2}\\
    &=-il_B^2,\\
\end{aligned}
\end{equation}
where we have used $\comm{r_\mu}{\pi_\nu} = \comm{r_\mu}{p_\nu}$. 

Eq. \ref{eq:R_x_R_y_comm} allows us to bound the degree to which $R_x$ and $R_y$ may be simultaneously localized. Letting $\sigma(R_\mu)$ be the standard deviation of $R_\mu$, the AM-GM inequality implies that:
\begin{equation}\label{eq:AM_GM}
    \sigma(\v R) = \sqrt{\sigma(R_x)^2 + \sigma(R_y)^2} \geq \sqrt{2 \sigma(R_x) \sigma(R_y)}.
\end{equation}
The Heisenberg uncertainty principle requires that:
\begin{equation}\label{eq:Heisenberg}
    \sigma(R_x) \sigma(R_y) \geq \frac{1}{2}|\expval{\comm{R_x}{R_y}}| = \frac{l_B^2}{2}.
\end{equation}
Thus, we have:
\begin{equation}\label{eq:R_uncertainty}
    \sigma(\v R) \geq l_B,
\end{equation}
or in other words, the guiding center may not be localized beyond the magnetic length.

Our next goal is to construct magnetic translation operators $T(\v a)$ such that $T^\dagger(\v a) \v R T(\v a) = \v R + \v a$. The commutator between $R_x$ and $R_y$ suggests the following guess for a translation operator:
\begin{equation}\label{eq:T_guess}
    T(\v a) = \exp(-i\frac{\epsilon_{\mu\nu}}{l_B^2}R_\mu a_\nu).
\end{equation}
In the following, two identities will be helpful. Assuming that $A$ and $B$ are linear operators such that $\comm{A}{B}$ commutes with $A$ and $B$ (this is true for us, because $R_x$ and $R_y$'s commutator is just a complex number):
\begin{equation}\label{eq:comm_identities_1}
\begin{aligned}
    e^A B e^{-A} &= B + \comm{A}{B}, \\ 
    e^A e^B &= e^{A + B + \frac{1}{2}\comm{A}{B}}.
\end{aligned}
\end{equation}
Then we have:
\begin{equation}\label{eq:trans_action}
\begin{aligned}
    &T^\dagger(\v a) R_\kappa T(\v a) \\
    =& \exp(+i\frac{\epsilon_{\mu\nu}}{l_B^2}R_\mu a_\nu) R_\kappa \exp(-i\frac{\epsilon_{\mu\nu}}{l_B^2}R_\mu a_\nu)\\
    =& R_\kappa +i\frac{\epsilon_{\mu\nu}}{l_B^2}a_\nu\comm{R_\mu}{R_\kappa}\\
    =& R_\kappa +i\frac{\epsilon_{\mu\nu}}{l_B^2}a_\nu (-i l_B^2 \epsilon_{\mu\kappa})\\
    =& (R + a)_k,
\end{aligned}
\end{equation}
showing that the proposed operator performs the translation as desired.

An important feature of the magnetic translation operators is that they commute only up to a phase. We have:
\begin{equation}\label{eq:trans_comm}
\begin{aligned}
    &T(\v a) T(\v b) \\
    =& \exp(\frac{1}{2}\comm{-i\frac{\epsilon_{\mu\nu}}{l_B^2}R_\mu a_\nu}{-i\frac{\epsilon_{\kappa\lambda}}{l_B^2}R_\kappa b_\lambda}) T(\v a + \v b)\\
    =& \exp(-\frac{\epsilon_{\mu\nu} \epsilon_{\kappa\lambda}}{2l_B^4} a_\nu b_\lambda \comm{R_\mu}{R_k}) T(\v a + \v b)\\
    =& \exp(-\frac{\epsilon_{\mu\nu} \epsilon_{\kappa\lambda}}{2l_B^4} a_\nu b_\lambda (-i\epsilon_{\mu\kappa}l_B^2)) T(\v a + \v b)\\
    =& \exp(i\frac{\epsilon_{\nu\lambda}}{2l_B^2} a_\nu b_\lambda) T(\v a + \v b)\\
    =& \exp(i\frac{\v a \wedge \v b}{2l_B^2}) T(\v a + \v b),\\
\end{aligned}
\end{equation}
where $\v a \wedge \v b = a_x b_y - a_y b_x$ is the two-dimensional wedge product. Eq. \ref{eq:trans_comm} yields the product of four translation operators around a loop:
\begin{equation}\label{eq:trans_loop}
    T(\v a) T(\v b) T(-\v a) T(-\v b) = \exp(i\frac{\v a \wedge \v b}{l_B^2}),
\end{equation}
which is consistent with the Aharanov-Bohm phase accumulated over such a path.

\subsection{Symmetric Gauge Operators}

For our system, it is convenient to work in the symmetric gauge $\v A = \v B \cross \v r / 2 = \left(-By/2, Bx/2, 0\right)$, which manifestly respects rotation symmetry. In the symmetric gauge, the natural coordinates are the complex $z = x + iy$ and its conjugate $z^* = x - iy$. The derivative conversions are:
\begin{equation}\label{eq:deriv_conv_forward}
\begin{aligned}
    \partial &\equiv \pdv{}{z} = \pdv{}{x}\pdv{x}{z} + \pdv{}{y}\pdv{y}{z} = \frac{1}{2}\left(\pdv{}{x} -i \pdv{}{y}\right),\\
    \bar{\partial} &\equiv \pdv{}{\bar{z}} = \pdv{}{x}\pdv{x}{\bar{z}} + \pdv{}{y}\pdv{y}{\bar{z}} = \frac{1}{2}\left(\pdv{}{x} +i \pdv{}{y}\right),\\
    \pdv{}{x} &= \pdv{}{z}\pdv{z}{x} + \pdv{}{\bar{z}}\pdv{\bar{z}}{x} = \partial + \bar{\partial},\\
    \pdv{}{y} &= \pdv{}{z}\pdv{z}{y} + \pdv{}{\bar{z}}\pdv{\bar{z}}{y} = i(\partial - \bar{\partial}),\\
\end{aligned}
\end{equation}
where $\partial$ and $\bar{\partial}$ treat $z$ and $\bar{z}$ as independent. Because $\left(\partial_x\right)^\dagger = - \partial_x$ and $\left(\partial_y\right)^\dagger = - \partial_y$, we have $\partial^\dagger = - \bar{\partial}$. It is also helpful to remember that $A_x + i A_y = iBz/2$ and $A_x - i A_y = -iB\bar{z}/2$.
In terms of the complex coordinates, the Landau level raising and lowering operators become:
\begin{equation}\label{eq:a_complex}
\begin{aligned}
    a &= \frac{l_B}{\sqrt{2}}\left( -i \partial_x - q A_x + \partial_y - iq A_y \right)\\
    &= \frac{l_B}{\sqrt{2}}\left( -2i \bar{\partial} - \frac{iqBz}{2}\right)\\
     &= -i\sqrt{2} \left(l_B \bar{\partial} + \frac{z}{4l_B}\right),\\
    a^\dagger &= -i\sqrt{2} \left(l_B \partial - \frac{\bar{z}}{4l_B} \right).\\
\end{aligned}
\end{equation}

The (canonical) angular momentum is defined as:
\begin{equation}\label{eq:Lz_def}
    L = x p_y - y p_x,
\end{equation}
which may be rewritten in terms of the complex coordinates as:
\begin{equation}\label{eq:Lz_complex}
\begin{aligned}
    L &= -i\left(\frac{z + \bar{z}}{2}\right) \cdot i\left(\partial - \bar{\partial}\right) +i\left(\frac{z - \bar{z}}{2i}\right) \cdot \left(\partial + \bar{\partial} \right)\\
    &= z \partial - \bar{z}\bar{\partial}.\\
\end{aligned}
\end{equation}
Acting on a term, $L$ counts the number of $z$'s minus the number of $\bar{z}$'s, which makes sense as each $z$ ($\bar{z}$) contributes a factor of $+1$ ($-1$) to the winding about the origin.

We desire operators that simultaneously raise or lower $L$ while commuting with the Hamiltonian. The previously defined guiding center operators provide a natural way to do this, given that they commute with the Hamiltonian and may be combined in a way like $R_x + i R_y$ to increase the wavefunction's winding about the origin. Consider:
\begin{equation}\label{eq:b_def}
\begin{aligned}
    b &= \frac{-i}{\sqrt{2}l_B}\left(R_x - i R_y\right),\\
    b^\dagger &= \frac{i}{\sqrt{2}l_B}\left(R_x + i R_y\right),
\end{aligned}
\end{equation}
which in terms of the complex coordinates is:
\begin{equation}\label{eq:b_complex}
\begin{aligned}
    b &= \frac{-i}{\sqrt{2}l_B}\left( x + l_B^2 \pi_y - iy + i l_B^2 \pi_x \right)\\
      &= \frac{-i}{\sqrt{2}l_B}\left( \bar{z} + il_B^2 \left[\pi_x - i\pi_y\right] \right)\\
      &= \frac{-i}{\sqrt{2}l_B}\left( \bar{z} + il_B^2 \left[ -2i\partial + \frac{iqB\bar{z}}{2} \right] \right)\\
     &= -i\sqrt{2} \left(l_B \partial + \frac{\bar{z}}{4l_B}\right),\\
    b^\dagger &= -i\sqrt{2} \left(l_B \bar{\partial} - \frac{z}{4l_B} \right).\\
\end{aligned}
\end{equation}
Because $b$ either kills a factor of $z$ or creates a factor of $\bar{z}$ when acting on a state, it is clear that $b$ lowers the angular momentum by one, and likewise $b^\dagger$ clearly acts as a raising operator.

It is helpful for later derivations to express $z$, $\bar{z}$, $\partial$, $\bar{\partial}$ in terms of the $a$ and $b$ operators. From Eqs. \ref{eq:a_complex} and \ref{eq:b_complex}, a short derivation shows that:
\begin{equation}\label{eq:z_to_ab}
\begin{aligned}
    z &= i\sqrt{2}l_B \left( a - b^\dagger \right),\\
    \bar{z} &= -i\sqrt{2} l_B \left( a^\dagger - b \right),\\
    \partial &= \frac{i}{2\sqrt{2} l_B} \left( a^\dagger + b \right),\\
    \bar{\partial} &= \frac{i}{2\sqrt{2} l_B} \left( a + b^\dagger \right).\\
\end{aligned}
\end{equation}
Similarly, the angular momentum may be expressed as:
\begin{equation}\label{eq:L_ab}
    L = -a^\dagger a + b^\dagger b.
\end{equation}

For a many-particle system, the total angular momentum is:
\begin{equation}\label{eq:L_tot_MB}
    L_\text{tot} = \sum_{i=1}^N L_i = \sum_{i=1}^N \left(-a^\dagger_i a_i + b^\dagger_i b_i\right),
\end{equation}
where $a_i$ denotes the Landau level lowering operator for particle $i$, and likewise for $b_i$. The many-body Hamiltonian is:
\begin{equation}\label{eq:H_ab_exp}
\begin{aligned}
    H &= T + V,\\
    T &= \omega_B \sum_i \left( a_i^\dagger a_i + \frac{1}{2} \right),\\
    V &= \sum_{i < j} v\left( \sqrt{\left[ \bar{z}_i - \bar{z}_j \right] \left[ z_i - z_j \right]} \right).
\end{aligned}
\end{equation}
$L_\text{tot}$ commutes with $T$ because $L_\text{tot}$ preserves the total number of $a$-quanta. To see that $L_\text{tot}$ commutes with $V$, first note that any complex coordinate $z_i$ increments $L_\text{tot}$ by one (from Eq. \ref{eq:z_to_ab}, $z_i$ either creates a $b$-quanta or destroys an $a$-quanta), while any conjugate coordinate $\bar{z}_i$ decrements $L_\text{tot}$, so any operator such as $V$ with a balanced number of $z$'s and $\bar{z}$'s commutes with $L_\text{tot}$. Therefore, the total angular momentum is conserved quantity.

\subsection{Center-of-Mass Angular Momentum}

When there is no external confining potential, there is an additional conserved quantity: the center-of-mass angular momentum. Intuitively, this is conserved because all electrons can be displaced simultaneously without changing the energy. To show this mathematically, we transform into the center-of-mass frame, where we have the center-of-mass coordinate:
\begin{equation}\label{eq:z_COM_def}
    Z_1 = \frac{1}{N}\sum_i z_i,
\end{equation}
along with $N-1$ relative coordinates such as $z_2 - z_1, z_3 - z_1, \ldots, z_N - z_1$. Following \ref{eq:Lz_def}, we may define the center-of-mass angular momentum operator:
\begin{equation}\label{eq:Lz_COM_def}
    L_\text{CM} = Z_1\pdv{}{Z_1} - \bar{Z}_1\pdv{}{\bar{Z}_1}.
\end{equation}
If one moves only the center-of-mass by a displacement $\Delta z$ but does not alter the $N - 1$ relative coordinates, then each of the $z_i$ must also displace by $\Delta z$. Therefore, we have $\pdv{z_i}{Z_1} = 1$, allowing us to re-express Eq. \ref{eq:Lz_COM_def} as:
\begin{equation}\label{eq:Lz_COM_two}
\begin{aligned}
    L_\text{CM} &= \frac{1}{N}\sum_i \sum_j \left[ z_i\pdv{}{z_j} \pdv{z_j}{Z_1} - \bar{z}_i\pdv{}{\bar{z}_j} \pdv{\bar{z}_j}{\bar{Z}_1} \right]\\
    &= \frac{1}{N}\sum_{ij} \left(z_i \partial_j - \bar{z}_i \bar{\partial}_j\right).
\end{aligned}
\end{equation}

In terms of the $a$ and $b$ ladder operators, the center-of-mass angular momentum is:
\begin{equation}\label{eq:Lz_COM_ab}
\begin{aligned}
    L_\text{CM} =& \frac{1}{N}\sum_{ij} \left(z_i \partial_j - \bar{z}_i \bar{\partial}_j\right)\\
    =& \frac{1}{N}\sum_{ij} \bigg[-\frac{1}{2} \left( a_i - b^\dagger_i \right) \left( a^\dagger_j + b_j \right)\\
    & -\frac{1}{2} \left( a^\dagger_i - b_i \right) \left( a_j + b^\dagger_j \right)\bigg]\\
    =&\frac{1}{N} \sum_{ij}\left[- a_i^\dagger a_j + b_i^\dagger b_j \right].\\
\end{aligned}
\end{equation}
We may express the center-of-mass angular momentum more concisely in terms of ``center-of-mass ladder operators'':
\begin{equation}\label{eq:COM_ladder}
\begin{aligned}
    a_\text{CM} &= \frac{a_1 + a_2 + \ldots + a_N}{\sqrt{N}},\\
    b_\text{CM} &= \frac{b_1 + b_2 + \ldots + b_N}{\sqrt{N}},\\
    L_\text{CM} &= -a_\text{CM}^\dagger a_\text{CM} + b_\text{CM}^\dagger b_\text{CM}.
\end{aligned}
\end{equation}

From Eq. \ref{eq:Lz_COM_ab}, it is clear that $L_\text{CM}$ does not alter the number of a $a$ or $b$-quanta, so $L_\text{CM}$ commutes with $L_\text{tot}$. Due to to translation invariance, $V$ only depends on the differences between coordinates and thus can be expressed using only differences between ladder operators, such as $a_{ij} \equiv a_i - a_j$ or $b_{ij} \equiv b_i - b_j$. It is easy to see that the difference operators $a_{ij}$ and $b_{ij}$ must always commute with the center-of-mass operators (which is an equal linear combination of all the single-particle operators), so $a_\text{CM}$ and $b_\text{CM}$ commute with $V$. $a_\text{CM}$ does not commute with the kinetic energy $T$, but $a_\text{CM}^\dagger a_\text{CM}$ does. Therefore, the ``center-of-mass Landau level'' $a_\text{CM}^\dagger a_\text{CM}$ and center-of-mass angular momentum $-a_\text{CM}^\dagger a_\text{CM} + b_\text{CM}^\dagger b_\text{CM}$ index symmetry sectors. Without Landau level mixing, $a^\dagger_\text{CM} a_\text{CM} = 0$ and is irrelevant.

\section{Exact Diagonalization Matrix Elements}

\subsection{Single-Particle Basis}

The unique normalized state satisfying $a\ket{n} = 0$ and $b^\dagger b \ket{n} = n\ket{n}$ is:
\begin{equation}\label{eq:one_body_eigenstate}
    \ket{n} = \frac{1}{\sqrt{2\pi n! \, 2^n}}z^ne^{-\frac{|z|^2}{4}},
\end{equation}
where we take $l_B = 1$ from here on for convenience. To see this rigorously, consider a wavefunction $\psi(z, \bar{z}) = f(z, \bar{z}) \exp(-|z|^2/4)$ (one may always choose to parameterize a wavefunction in this form) that is annihilated by $a$:
\begin{equation}\label{eq:sym_gauge_LLL}
\begin{aligned}
    -i\sqrt{2} \left(\bar{\partial} + \frac{z}{4}\right)\left(f(z, \bar{z}) e^{-\frac{|z|^2}{4}}\right) &= 0\\
    \left[\bar{\partial}f(z, \bar{z})\right] e^{-\frac{|z|^2}{4}} - f(z, \bar{z}) \frac{z}{4} e^{-\frac{|z|^2}{4}} &\\
    +\frac{z}{4} \left(f(z, \bar{z}) e^{-\frac{|z|^2}{4}}\right) &= 0\\
    \bar{\partial}f(z, \bar{z}) &= 0.\\
\end{aligned}
\end{equation}
Thus, any wavefunction lying entirely in the LLL must be the product of a holomorphic function $f(z)$ and a Gaussian envelope $\exp(-|z|^2/4)$. Demanding that we have an angular momentum eigenstate forces $f(z)$ to to be of constant degree $f(z) \propto z^n$, and the normalization in Eq. \ref{eq:one_body_eigenstate} is easily determined from standard integrals.

\subsection{Two-Body Matrix Elements}

To build the many-body Hamiltonian, we need the matrix elements between two-body product states:
\begin{equation}\label{eq:matrix_elem_def}
    V^l_{n_1 n_2} = \bra{n_1 + l, n_2}v(|z_1 - z_2|)\ket{n_1, n_2 + l},
\end{equation}
where the two-body product states are:
\begin{equation}\label{eq:two_body_product}
    \ket{n_1, n_2} = \frac{1}{2\pi\sqrt{n_1! \, n_2! \, 2^{n_1 + n_2}}}z_1^{n_1} z_2^{n_2} e^{-\frac{|z_1|^2 + |z_2|^2}{4}}.
\end{equation}
Here, we demonstrate a method for analytically calculating the matrix elements of any interaction in terms of the corresponding Haldane pseudopotential and hypergeometric functions. 

To start, observe that the two-body problem in the LLL is exactly solvable because both the center-of-mass angular momentum and relative angular momentum are conserved by the electron-electron interaction. For a prescribed center-of-mass angular momentum $M$ and relative angular momentum $m$, the only possible state is:
\begin{equation}\label{eq:COM_basis}
\begin{gathered}
    v = (z_1 + z_2)/\sqrt{2}, \quad w = (z_1 - z_2) / \sqrt{2},\\
    \ket{M, m}_\text{rel} = \frac{1}{2\pi\sqrt{M! \, m! \, 2^{M + m}}}v^M w^m e^{-\frac{|v|^2 + |w|^2}{4}}.
\end{gathered}
\end{equation}
The corresponding eigenvalue is independent of $M$ and is known as the Haldane pseudopotential:
\begin{equation}\label{eq:pseudopotential_formula}
\begin{aligned}
    v_m &= \bra{M, m}_\text{rel} v(|z_1 - z_2|) \ket{M, m}_\text{rel} \\
    &= \frac{1}{2\pi m! \, 2^m} \int \text{d}^2w \left[v\left(\sqrt{2}|w|\right) |w|^{2m} e^{-\frac{|w|^2}{2}}\right]\\
    &= \frac{1}{m!} \int_{u = \frac{w^2}{2} = 0}^{u = \infty} \text{d}u \left[v\left(2 \sqrt{u}\right)u^m e^{-u}\right].
\end{aligned}
\end{equation}

In general, $u^m e^{-u}$ is peaked near $u \approx m$, so $v_m \sim v(2\sqrt{m})$ for large $m$. This is expected, as the amplitude of $\ket{M, m}_\text{rel}$ is peaked when the electrons are a distance of $2\sqrt{m}$ apart. For bosons, the minimum relative angular momentum channel is $m = 0$, so for an interaction to be well-defined, it must diverge slower than $v(r) \propto 1/r^2$. For fermions, the minimum relative angular momentum channel is $m = 1$, so for an interaction to be well-defined, it must diverge slower than $v(r) \propto 1/r^4$.

Using the definition of the gamma function, Eq. \ref{eq:pseudopotential_formula} can be evaluated analytically for power-law interactions $v(r) = r^\alpha$:
\begin{equation}\label{eq:power_law_pseudopotential}
    v_m\left[ v(r) = r^\alpha \right] =\frac{2^\alpha \Gamma(m + 1 + \frac{\alpha}{2})}{\Gamma(m + 1)}.
\end{equation}
We may also consider logarithmic repulsion $v(r) = -\ln(r)$, for which the pseudopotential becomes:
\begin{equation}\label{eq:logarithmic_pseudopotential}
\begin{aligned}
    &v_m\left[ v(r) = -\ln(r) \right] \\
    =& -\frac{1}{m!} \int_{u = \frac{w^2}{2} = 0}^{u = \infty} \text{d}u \left[\ln \left(2 \sqrt{u}\right)u^m e^{-u}\right]\\
    =& -\ln(2) - \frac{1}{2\Gamma(m + 1)} \int_{u = 0}^{u = \infty} \text{d}u \left[\ln \left( u \right)u^m e^{-u}\right]\\
    =& -\ln(2) - \frac{1}{2\Gamma(m + 1)} \dv{}{m}\left( \int_{u = 0}^{u = \infty} \text{d}u \left[u^m e^{-u}\right] \right)\\
    =& -\ln(2) - \frac{1}{2} \psi(m + 1), \quad \psi(m) = \dv{}{m} \ln \Gamma(m).\\
\end{aligned}
\end{equation}
Above, $\psi(m)$ is known as the digamma function and may be computed using standard libraries.

Then to calculate $V^l_{n_1 n_2}$, all that we need to do is express $\ket{n_1, n_2}$ as a linear combination of the center-of-mass and relative angular momentum eigenstates $\ket{M, m}_\text{rel}$, in which the interaction is diagonal. This expansion may be preformed by substituting $z_1 = (v + w)/\sqrt{2}, z_2 = (v - w)/\sqrt{2}$ and expanding out using the binomial theorem:
\begin{widetext}
\begin{equation}\label{eq:COM_expansion_1}
\begin{aligned}
    \ket{n_1, n_2} &= \frac{1}{2\pi\sqrt{n_1! \, n_2! \, 2^{n_1 + n_2}}}\bigg(\frac{v + w}{\sqrt{2}}\bigg)^{n_1} \bigg(\frac{v - w}{\sqrt{2}}\bigg)^{n_2} e^{-\frac{|v|^2 + |w|^2}{4}}\\
    &= \frac{1}{2\pi\sqrt{n_1! \, n_2!} \, 2^{n_1 + n_2}}(v + w)^{n_1} (v - w)^{n_2} e^{-\frac{|v|^2 + |w|^2}{4}}\\
    &= \frac{1}{2\pi\sqrt{n_1! \, n_2!} \, 2^{n_1 + n_2}}\Bigg[\sum_{a=0}^{n_1} \binom{n_1}{a} v^{n_1 - a}w^a \Bigg] \Bigg[\sum_{b=0}^{n_2} (-1)^b \binom{n_2}{b} v^{n_2 - b}w^b\Bigg] e^{-\frac{|v|^2 + |w|^2}{4}}\\
    &= \frac{1}{2\pi\sqrt{n_1! \, n_2!} \, 2^{n_1 + n_2}} \sum_{m=0}^{n_1 + n_2}\left[\sum_{\substack{0 \leq a \leq n_1 \\ 0 \leq b \leq n_2 \\ a + b = m}} (-1)^b \binom{n_1}{a} \binom{n_2}{b} \right]v^{n_1 + n_2 - m}w^m e^{-\frac{|v|^2 + |w|^2}{4}}\\
    &= \sum_{m=0}^{n_1 + n_2}\left[\frac{2\pi\sqrt{(n_1 + n_2 - m)! \, m! \, 2^{n_1 + n_2}}}{2\pi\sqrt{n_1! \, n_2!} \, 2^{n_1 + n_2}} \sum_{\substack{0 \leq a \leq n_1 \\ 0 \leq b \leq n_2 \\ a + b = m}} (-1)^b \binom{n_1}{a} \binom{n_2}{b} \right] \ket{n_1 + n_2 - m, m}_\text{rel}\\
    &= \sum_{m=0}^{n_1 + n_2}\left[\sqrt{\frac{(n_1 + n_2 - m)! \, m!}{n_1! \, n_2! \, 2^{n_1 + n_2}}} \sum_{a = \max(m-n_2, 0)}^{\min(m, n_1)} (-1)^{m-a} \binom{n_1}{a} \binom{n_2}{m - a} \right] \ket{n_1 + n_2 - m, m}_\text{rel}.
\end{aligned}
\end{equation}
The combinatorial sum over $a$ becomes difficult to evaluate numerically for high $n_1$ and $n_2$ due to the alternating sign $(-1)^{m-a}$. Luckily, the sum may be expressed in terms of a hypergeometric function:
\begin{equation}\label{eq:combinatorial_part}
    \sum_{a = \max(m-n_2, 0)}^{\min(m, n_1)} (-1)^{m-a} \binom{n_1}{a} \binom{n_2}{m - a} = 
    \begin{dcases} 
      (-1)^m \binom{n_2}{m} {}_2F_1(-m, -n_1; 1 - m + n_2; -1) & m \leq n_2 \\
      (-1)^{n_2} \binom{n_1}{m - n_2} {}_2F_1(m - n_1 - n_2, -n_2; 1 + m - n_2; -1) & m > n_2 \\
   \end{dcases}.
\end{equation}
Here, ${}_2F_1$ is the Gauss hypergeometric function expressed as the following power series:
\begin{equation}\label{eq:2F1_def}
    {}_2F_1(a,b;c;z) = \sum_{n=0}^\infty \frac{(a)_n (b)_n}{(c)_n} \frac{z^n}{n!}, \quad
     (q)_n = \begin{cases}  1  & n = 0 \\
    q(q+1) \cdots (q+n-1) & n > 0
\end{cases}.
\end{equation}
If one of the first two arguments is nonpositive, then the power series is finite:
\begin{equation}\label{eq:2F1_finite}
    {}_2F_1(-m,b;c;z) = \sum_{n=0}^m (-1)^n \binom{m}{n} \frac{(b)_n}{(c)_n} z^n,
\end{equation}
where $m$ is assumed positive. Eq. \ref{eq:2F1_finite} can be used to prove the correctness of Eq. \ref{eq:combinatorial_part}. For the first case, where $m \leq n_2$, both the first and second argument of ${}_2F_1$ are nonpositive:
\begin{equation}\label{eq:combinatorial_first_case}
\begin{aligned}
    (-1)^m \binom{n_2}{m} {}_2F_1(-m, -n_1; 1 - m + n_2; -1) &= (-1)^m \binom{n_2}{m} \sum_{a=0}^{m} (-1)^a \binom{m}{a}\frac{(-n_1)_a}{(1 - m + n_2)_a}(-1)^a\\
    &= (-1)^m \frac{n_2!}{m! \, (n_2 - m)!} \sum_{a=0}^{\min(m, n_1)} \frac{m!}{a! \, (m-a)!}\frac{(-1)^a \frac{n_1!}{(n_1 - a)!}}{\frac{(1 - m + n_2 + a - 1)!}{(1 - m + n_2 - 1)!}}\\
    &= \sum_{a=\max(m - n_2, 0)}^{\min(m, n_1)} (-1)^{m - a} \binom{n_1}{a} \binom{n_2}{m - a}.\\
\end{aligned}
\end{equation}
For the second case, where $m > n_2$:
\begin{equation}
\begin{aligned}
    &(-1)^{n_2} \binom{n_1}{m - n_2} {}_2F_1(m - n_1 - n_2, -n_2; 1 + m - n_2; -1)\\
    =& (-1)^{n_2} \binom{n_1}{m - n_2} \sum_{b' = 0}^{\min(n_2, n_1 + n_2 - m)} (-1)^{b'} \binom{n_2}{b'} \frac{(m - n_1 - n_2)_{b'}}{(1 + m - n_2)_{b'}} (-1)^{b'}\\
    =& (-1)^{n_2} \frac{n_1 !}{(m - n_2)! \, (n_1 + n_2 - m)!} \sum_{b' = 0}^{n_2 + \min(0, n_1 - m)} \frac{n_2!}{b'! \, (n_2 - b')!} \frac{(-1)^{b'}\frac{(n_1 + n_2 - m)!}{(n_1 + n_2 - m - b')!}}{\frac{(1 + m - n_2 + b' - 1)!}{(1 + m - n_2 - 1)!}}\\
    =& \sum_{b' = 0}^{n_2 + \min(0, n_1 - m)} (-1)^{n_2 - b'} \binom{n_1}{m - n_2 + b'} \binom{n_2}{n_2 - b'}\\
    =& \sum_{a = m - n_2 + b' = \max(m - n_2, 0)}^{\min(m, n_1)} (-1)^{m - a} \binom{n_1}{a} \binom{n_2}{m - a}.\\
\end{aligned}
\end{equation}
Although it may appear that we have just wrapped a difficult sum into a convenient special function, there exist sophisticated algorithms for accurately calculating the hypergeometric function. In contrast, the original direct summation in Eq. \ref{eq:COM_expansion_1} is highly susceptible to precision loss and becomes unusable when $n_1$ and $n_2$ exceed 50.

Combining Eq. \ref{eq:COM_expansion_1} with Eq. \ref{eq:combinatorial_part} yields:
\begin{equation}\label{eq:final_COM_expansion}
 \begin{gathered}
    \ket{n_1, n_2} = \sum_{m = 0}^{n_1 + n_2} C^m_{n_1 n_2} \ket{n_1 + n_2 - m, m}_\text{rel},\\
%\end{equation}
%where the expansion coefficients $C^m_{n_1 n_2}$ are:
%\begin{equation}\label{eq:expansion_coefficients}
    C^m_{n_1 n_2} = \begin{dcases} 
      (-1)^m \sqrt{\frac{1}{2^{n_1 + n_2}} \binom{n_1 + n_2 - m}{n_1} \binom{n_2}{m}} {}_2F_1(-m, -n_1; 1 - m + n_2; -1)& m \leq n_2 \\
      (-1)^{n_2} \sqrt{\frac{1}{2^{n_1 + n_2}}\binom{m}{n_2} \binom{n_1}{m - n_2}} {}_2F_1(m - n_1 - n_2, -n_2; 1 + m - n_2; -1) & m > n_2 \\
   \end{dcases}.\\
\end{gathered}
\end{equation}
Then the Coulomb matrix elements may be calculated easily as:
\begin{equation}
    V^l_{n_1 n_2} = \bra{n_1 + l, n_2}v(|\v r_1 - \v r_2|)\ket{n_1, n_2 + l} = \sum_{m = 0}^{n_1 + n_2 + l} v_m \left(C^m_{n_1 + l, n_2}\right)^* C^m_{n_1, n_2 + l},
\end{equation}
\end{widetext}
where all terms in the sum are $O(1)$ because the expansion coefficients $C^m_{n_1 n_2}$ must satisfy the normalization condition $\sum_{m=0}^{n_1 + n_2} |C^m_{n_1 n_2}|^2$. 

Checking the orthogonality of different pairs of $\ket{n_1, n_2}$ provides a convenient way to detect numerical error. In our code, no precision loss beyond machine precision is observed even as $n_1$ and $n_2$ approach $100$. We note that specifically for the Coulomb interaction $v(r) = 1 /r$, there exist other numerically stable ways of calculating the two-body matrix elements \cite{Tsiper_2002}. We have verified that our method is consistent with others in the literature up to machine precision.

\subsection{Many-Body Matrix Elements}

For $N$ electrons, the many-body states are Slater determinants of symmetric gauge orbitals, which are stored in the code as a sorted list of occupied orbitals $\ket{i_1 < i_2 < \ldots < i_N}$. Throughout, we use capital Latin letters $I, J, \ldots$ to represent Slater determinants. In our exact diagonalization calculations, we work within a given total angular momentum sector, so the sum $i_1 + i_2 + \ldots + i_N = L_\text{tot}$ for each Slater determinant is constrained. We note that total angular momentum conservation means that the angular momentum of any individual electron is bounded: the maximum angular momentum that any electron may have is $l_\text{max} = L_\text{tot} - (0 + 1 + \ldots + N - 2) = L_\text{tot} - (N - 1)(N - 2)/2$, which is attained by the determinant $\ket{0, 1, 2, \ldots N - 2, l_\text{max}}$. This naturally renders both the single-particle basis and many-body Hilbert space finite. We note that one may not artificially cutoff the single-particle basis, as this is equivalent to enforcing a hard potential wall and thus spoils the center-of-mass angular momentum conservation.

Due to angular momentum conservation, a Slater determinant only has nonzero matrix element with itself or another Slater determinant connected by double replacement:
\begin{equation}\label{eq:Slater_condon}
\bra{J}H\ket{I} = 
\begin{dcases} 
    \frac{1}{2} \sum_{ij \in I} \bra{ij}v\bigg[\ket{ij} - \ket{ji}\bigg], & I = J \\
    \text{sgn}(\sigma) \bra{ab}v \bigg[\ket{ij} - \ket{ji} \bigg], & J = \sigma \left( I^{ab}_{ij} \right) \\
\end{dcases}.
\end{equation}
Above, $I^{ab}_{ij}$ denotes the Slater determinant constructed from replacing filled orbitals $i$ and $j$ with new orbitals $a$ and $b$, $\sigma$ is the permutation that sorts the resulting Slater determinant (permutations of the occupied orbitals do not generate distinct determinants), and $\text{sgn}(\sigma)$ is the sign of said permutation. In the second case, $i$, $j$, $a$, and $b$ must all be distinct.

To determine the action of the center-of-mass angular momentum lowering operator, we first put aside particle statistics temporarily. When working with $N$ distinguishable particles, a natural many-body state is the tensor product:
\begin{equation}\label{eq:boltzmanon}
    \ket{l_1, l_2, \ldots, l_N}_D \equiv \ket{l_1} \otimes \ket{l_2} \otimes \ldots \otimes \ket{l_N},
\end{equation}
where the ``$D$'' subscript on the LHS clarifies that the particles are ``distinguishable''. The action of the center-of-mass lowering operator on this state is:
\begin{equation}\label{eq:b_CM_dist_action}
\begin{aligned}
    b_\text{CM} \ket{l_1, l_2, \ldots, l_N}_D
    =&\sqrt{\frac{l_1}{N}} \ket{l_1 - 1, l_2, \ldots, l_N}_D\\
    &+\sqrt{\frac{l_2}{N}} \ket{l_1, l_2 - 1, \ldots, l_N}_D\\
    &\ldots \\
    &+\sqrt{\frac{l_N}{N}} \ket{l_1, l_2, \ldots, l_N - 1}_D.
\end{aligned}
\end{equation}
The antisymmetrization operator, which converts tensor product states $\ket{l_1, l_2, \ldots, l_N}_D$ into Slater determinants $\ket{l_1, l_2, \ldots, l_N}$, commutes with $b_\text{CM}$ since $b_\text{CM}$ treats all particles on equal footing. Thus, Eq. \ref{eq:b_CM_dist_action} applies similarly to Slater determinants:
\begin{equation}\label{eq:b_CM_ferm_action}
\begin{aligned}
    b_\text{CM} \ket{l_1, l_2, \ldots, l_N}
    =&\sqrt{\frac{l_1}{N}} \ket{l_1 - 1, l_2, \ldots, l_N}\\
    &+\sqrt{\frac{l_2}{N}} \ket{l_1, l_2 - 1, \ldots, l_N}\\
    &\ldots \\
    &+\sqrt{\frac{l_N}{N}} \ket{l_1, l_2, \ldots, l_N - 1},
\end{aligned}
\end{equation}
where we identify Slater determinants containing duplicated orbitals with the zero vector.

%\bibliography{SM_bib}

\end{document}